\documentclass{jfm}

\usepackage{graphicx}
\usepackage{natbib}

\ifCUPmtlplainloaded \else
  \checkfont{eurm10}
  \iffontfound
    \IfFileExists{upmath.sty}
      {\typeout{^^JFound AMS Euler Roman fonts on the system,
                   using the 'upmath' package.^^J}%
       \usepackage{upmath}}
      {\typeout{^^JFound AMS Euler Roman fonts on the system, but you
                   dont seem to have the}%
       \typeout{'upmath' package installed. JFM.cls can take advantage
                 of these fonts,^^Jif you use 'upmath' package.^^J}%
      }
  \else
  \fi
\fi

\ifCUPmtlplainloaded \else
  \checkfont{msam10}
  \iffontfound
    \IfFileExists{amssymb.sty}
      {\typeout{^^JFound AMS Symbol fonts on the system, using the
                'amssymb' package.^^J}%
       \usepackage{amssymb}%
         \let\leq=\leqslant
         
      }{}
  \fi
\fi

\ifCUPmtlplainloaded \else
  \IfFileExists{amsbsy.sty}
    {\typeout{^^JFound the 'amsbsy' package on the system, using it.^^J}%
     \usepackage{amsbsy}}
    {}
\fi

\newsavebox{\astrutbox}
\sbox{\astrutbox}{\rule[-5pt]{0pt}{20pt}}

\title[]{Transition to turbulence in duct flow}

\author[D. Biau, H. Soueid and A. Bottaro]%
{D\ls A\ls M\ls I\ls E\ls N\ns B\ls I\ls A\ls  U, \ns
H\ls O\ls U\ls S\ls S\ls A\ls M\ns S\ls  O\ls U\ls E\ls I\ls D\footnote{Also at the Institut de M\'{e}canique des Fluides de Toulouse, All\'{e}e du Pr. C. Soula, 31400 Toulouse, France}\break
\and A\ls L\ls E\ls S\ls S\ls A\ls N\ls D\ls R\ls O\ns B\ls O\ls T\ls
T\ls A\ls R\ls O}

\affiliation{%
Universit\`a di Genova, DICAT, Via Montallegro 1, 16145 Genova, Italy\\ }%

\pubyear{2008}
\volume{596}
\pagerange{133--142}
\date{11 July 2007 and in revised form 7 November 2007}
\begin{document}

\maketitle

\begin{abstract}
The transition of the flow in a duct of square cross-section is studied. Like in the similar case of the pipe flow, 
the motion is linearly stable for all Reynolds numbers; this flow is thus a good candidate to investigate the 'bypass' 
path to turbulence. Initially the so-called 'linear optimal perturbation problem' is formulated and solved, yielding
optimal disturbances in the form of longitudinal vortices. Such optimals, however, fail to
elicit a significant response from the system in the nonlinear regime. Thus, streamwise-inhomogeneous, sub-optimal 
disturbances are focussed upon; nonlinear quadratic interactions are immediately evoked by such initial perturbations 
and an unstable streamwise-homogeneous large amplitude mode rapidly emerges. The subsequent evolution of the flow, 
at a value of the Reynolds number at the edge between fully developed turbulence and relaminarization, shows the 
alternance of patterns with two pairs of large scale vortices near opposing parallel walls. Such edge states bear 
a resemblance to optimal disturbances.
\end{abstract}

\section{\label{sec:intro} Introduction}
Transition to turbulence in ducts is still an unsolved issue despite the 120-plus years since the observations by Osborne Reynolds that led to the definition of a similarity parameter, the ratio of the viscous to the convective time scale, capable of broadly separating the cases where the flow state was laminar from those where turbulence prevailed. Recent years have seen a resurgence of interest in the topic, spurred by new developments in linear and nonlinear stability theories. As is now well known, classical small perturbation theory is uncapable to provide an explanation for the onset of transition in ducts and pipes (\cite{Gill1965, Salwenetal1980, TatsumiYoshimura1990}). Current understanding ascribes the failure of classical theory to its focus on the asymptotic behavior of individual modes; when a small disturbance composed by a weighted combination of linear eigenfunctions is considered, there is the potential for very large short-time amplification of perturbation energy, even in nominally stable flow conditions.  This behavior has been reported by \cite{Landahl1980} and \cite{BobergBrosa1988} and has been given the name of algebraic instability (and later 'transient growth theory'), since the initial rapid growth in time of small disturbances goes like $t$. The property is related to the non-normality of the linearized stability operator (which does not commute with its adjoint).

Despite the appeal and elegance of transient growth theory, it was realized that the fully nonlinear Navier-Stokes equations need to be used to understand transition phenomena. In this context, we mention the work of \cite{Nagata1990, Nagata1997}, \cite{Waleffe1997, Waleffe1998, Waleffe2003},   \cite{FaisstEckhardt2003} and \cite{WedinKerswell2004}. These works present traveling wave and equilibrium solutions of the Navier-Stokes equation for channel and pipe flows that are possibly related to transition. Experimental investigations along these lines are due to \cite{Hofetal2004, Hofetal2005}. 


In dynamical systems' terminology it is argued that unstable travelling waves appear through saddle node bifurcations in phase space; the travelling waves found so far are all saddle points with low dimensional unstable manifolds. The saddles act by attracting the flow from the vicinity of the laminar state, and then repelling it away. For transitional or turbulent, yet moderate, values of $Re$ the flow wanders in phase space between few repelling states, spending much time in their vicinities, before being abruptly ejected away, so that experimental observations yield recurrent sequences of familiar patterns (\cite{Artuso, Kerswell2005}). 

A yet unanswered issue concerns the initial conditions that are most suited to yield such unstable states. Traditional emphasis on so-called \emph{optimal perturbations} may be misplaced. In fact, there is but a weak connection between the flow structures that grow the most during the linear transient phase and the chaotic flows found at large times. Such a connection for the case of the pipe flow concerns the so-called \emph{edge state} which sits on a separatrix between laminar and turbulent flows (\cite{Eckhardt2007, PringleKerswell}). This state, made up by two asymmetric vortices in the cross-section, resembles the optimal disturbance of transient growth theory (\cite{Bergstrom1993}). For the motion in a square duct there seems to be no connection at all: the low-$Re$ turbulent flow, when averaged in time and space, is characterized by eight secondary vortices symmetric about diagonals and bisection lines.  It seems reasonable to argue that such secondary structures represent the skeleton of the unstable periodic orbits, but the disturbances that grow the most in the linear transient phase are formed by two vortices, symmetric about a diagonal (\cite{GallettiBottaro2004}).  In both configurations, pipe and square-duct, the optimal perturbation is a stationary pseudo-mode, elongated in the streamwise direction, and not a travelling wave.  This is a generic occurrence in wall-bounded shear flows, and it does not bode well for the establishment of a simple, direct relation between small amplitude disturbances (excited in an initial receptivity phase) and finite amplitude wavelike states. Although nonlinear effects can be pinpointed right away as the culprit for the missing link between early stage of transition and late stages, there is scope for a receptivity analysis focussed on transiently growing initial conditions, followed by nonlinear simulations. A motivation for the search of wavelike structures during the early stages of transition is also provided by recent careful experiments (\cite{PeixinhoMullin2006}) on the reverse transition in pipe flow, where modulated wave trains are found to emerge from long-term transients.

The present paper starts by comparing the efficiency of optimal and sub-optimal perturbations in triggering transition to turbulence at a value of the Reynolds number $Re$ close to the threshold between laminar and turbulent flow; it further shows that the turbulent motion oscillates around edge states which display an intriguing resemblance to optimal disturbances, before relaminarization occurs.  Finally, an interpretation of the results is provided after projecting them onto a suitably defined phase space.

\section{Model configuration}\label{modconf}

The incompressible flow in a duct of square cross-section is an appealing configuration for the presence of geometrical symmetries capable to strongly constrain the patterns of motion. Countless studies have been devoted to the formation of secondary vortices in the turbulent regime (see \cite{Gavrilakis1992} for a direct numerical simulation approach) and, more recently, an attempt has been made to link the appearance of such large-scale coherent states to the vortices appearing during the initial, optimal transient phase of disturbance growth (\cite{GallettiBottaro2004, Bottaroetal2006}). The longitudinal laminar flow velocity component has an analytic form $U(y,z)$ available, for example, in \cite{TatsumiYoshimura1990}, with $y$ and $z$ cross-stream axes. After normalizing distances with the channel height $h$, velocities with the friction velocity $u_\tau$, with $u_\tau^2=-\frac{h}{4\rho} \frac{dP}{dx}$, time with $h/u_\tau$ and pressure with $\rho u_\tau^2$, the following equations are found to govern the behaviour of the developed flow in an infinite duct:
\begin{eqnarray*}
\begin{array}{ll}
u_x + v_y + w_z = 0,\\ 
u_t +uu_x + vu_y + wu_z = -p_x + \frac{1}{Re_\tau}\Delta u + 4,\\ 
v_t +uv_x + vv_y + wv_z = -p_y + \frac{1}{Re_\tau}\Delta v,\\ 
w_t +uw_x + vw_y + ww_z = -p_z + \frac{1}{Re_\tau}\Delta w,
\end{array}
\label{ndimomeg}
\end{eqnarray*}
with $\Delta=\partial_{xx}+\partial_{yy}+\partial_{zz}$ and $Re_\tau=u_\tau~h/\nu$. By using $u_\tau$ as velocity scale we fix the pressure gradient, rather than the flow rate.

An incompressible pseudo-spectral solver, based on Chebyshev collocation in $y$ and $z$ and Fourier transform along $x$, has been employed to solve these equations. For time-integration a third-order semi-implicit backward differentiation/Adams-Bashforth scheme is used. In order to compute a pressure unpolluted by spurious modes, the pressure is approximated by polynomials ($P_{N-2}$) of two units lower-order than for the velocity ($P_N$). Only one collocation grid is used, and no pressure boundary condition is needed. The accuracy and stability properties of the method are discussed by \cite{Botella}. An adequate grid at $Re_\tau=150$ has been found to be composed by $51 \times 51$ Chebyshev points, with $N_x=128$ streamwise grid points or $84$ Fourier modes after de-aliasing. The $x$-length of the domain has been chosen equal to $4 \pi$ to accommodate a sufficiently large range of wavenumbers $\alpha$, with periodic boundary conditions. A finer grid resolution has also been used for fully developed turbulent flow with $71 \times 71 \times 256$ physical grid points, or $71\times 71 \times 170$ spectral modes and a streamwise length $L_x=6\pi$\footnote{With this resolution we obtain an excellent match with the results by \cite{Gavrilakis1992} for $Re_\tau=300$.}. The finer resolution run provides a slightly larger value of the threshold energy for transition, but integral quantities such as disturbance energy or skin friction factor are only marginally affected.  Since the threshold value is, in any case, a function also of the shape of the initial condition, we do not deem indispensable to pursue expensive calculations to determine it exactly.  For all cases studied, an adequate time step has been found to be $\Delta t=5\times 10^{-4}$.  

\begin{figure}
\begin{center}
\includegraphics[width=7cm]{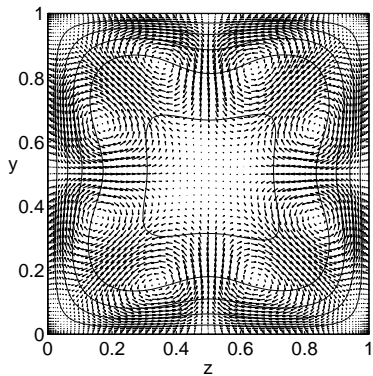}
\caption{\label{turb} Turbulent mean cross flow vortices and streamwise flow contours; isolines are spaced by $4~ u_\tau$.}
\end{center}
\end{figure}

The mean value of the generic function $g$ is defined as $\overline{g}(y,z) = \frac{1}{L_x~T}\int_{xt} g(x,y,z,t)~dxdt$. The addition of time averaging is necessary because of the finite (relatively low) streamwise length. The bulk velocity is $U_b=\int_{yz}\overline{u}~dydz$ and the centreline velocity is $U_c=\overline{u}(0.5,0.5)$. The friction factor for the square duct can be written as $f = 8~u_\tau^2 / U_b^2$. Some representative results are given in Table \ref{tab}, while the secondary flow field at $Re_\tau = 150$, averaged over forty units of time, is shown in figure \ref{turb}. It displays a very regular pattern with eight vortices, despite the fact that no averaging over quadrants has been performed.

\begin{table} 
  \begin{center} 
  \begin{tabular}{ccccc} 
          &   $f$  &  $U_c/U_b$  &  $Re_b$  &  $Re_c$  \\[3pt]  
laminar   & 0.018  &  2.0963     &  3163    &  6630.4  \\
turbulent & 0.0415 &  1.53       &  2084    &  3188 
  \end{tabular}
  \caption{\label{tab} Comparison of some numerical values for laminar and fully-developed turbulent flow at $Re_\tau=150$. The subscript $b$ refers to bulk and $c$ to centreline. The skin friction $f$ given by the empirical correlation by \cite{Jones} is $f=0.0481$.}
  \end{center} 
\end{table} 

\section{Optimal perturbations}

The Navier-Stokes equations, linearized around the ideal laminar flow, are:
 \begin{eqnarray*}\label{LST2dLNS}
i \alpha u +v_y+w_z & =&0, \\
u_t +i\alpha U u + vU_y+wU_z &=& -i\alpha p + (- \alpha^2 u + u_{yy}+u_{zz} )/Re_\tau,\\
v_t +i\alpha U v             &=& -      p_y + (- \alpha^2 v + v_{yy}+v_{zz} )/Re_\tau,\\
w_t +i\alpha U w             &=& -      p_z + (- \alpha^2 w + w_{yy}+w_{zz} )/Re_\tau,
\end{eqnarray*}
associated to boundary conditions $u=v=w=0$ on the walls; $\alpha$ is the streamwise wavenumber. The equations are integrated from a given initial condition at $t=0$ up to a final target time $t=T$. To identify the flow state at $t=0$ producing the largest disturbance growth at any given $T$, a variational technique, based on the repeated numerical integration of direct and adjoint stability equations, is used, coupled with transfer and optimality conditions (\cite{CorbettBottaro2000}). The functional for which optimization is sought is based on an energy-like norm and reads:
\begin{eqnarray*}
G(T)=\frac{E(T)}{E(0)},\quad \mbox{with} \quad E=\frac{1}{2} \int_y \int_z (u^* u + v^* v+ w^* w)\quad dy~dz,
\end{eqnarray*}
with $*$ denoting complex conjugate.


As a preliminary result the energy stability limit for this flow is determined; the minimal Reynolds number below which disturbances decrease monotonically is found to be $Re_\tau=23.22$ for $\alpha=0$. In terms of Reynolds number based on centreline velocity and half channel height, this limiting value is equal to $Re_c=79.44$. For comparison, in the case of Poiseuille flow, the Reynolds number is $Re_c\simeq49$, with $\alpha=0$ and spanwise wavenumber $\beta\simeq2$ (\emph{cf.} \cite{SchmidHenningson2000}).

\begin{figure}
\begin{center}
\includegraphics[width=9cm]{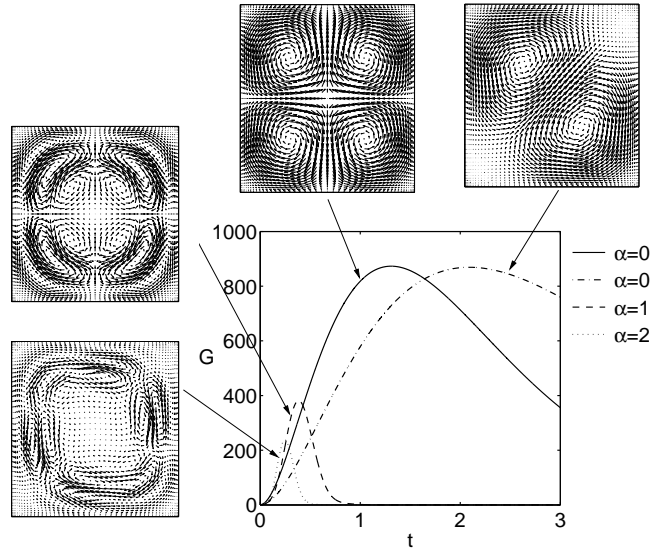}
\caption{\label{gain} 
Cross-flow velocity vectors for the optimal disturbances for $\alpha = 0, 1$ and $2$. Note that for $\alpha \ne 0$ the streamwise velocity fluctuations do not vanish at $t=0$.}
\end{center}
\end{figure}

Then, we compute optimal perturbations at $Re_\tau=150$, for 3 different Fourier modes: $\alpha=0,~ 1,~ 2$. The gain and the corresponding cross-flow optimal disturbances at $t=0$ are presented in figure \ref{gain}.
A global optimal solution ($G = 873.11$) is found for $\alpha=0$ at a time $T = 1.31$ and a comparable gain ($G=869.03$) is found at a later time ($T = 2.09$) for a solution which is topologically different. These two solutions, which we call 'global optimals' represent streamwise vortices of vanishing streamwise disturbance velocity; they evolve downstream producing streaks of high and low longitudinal velocity. A global optimal state with two cells arranged along a duct diagonal, was obtained by \cite{GallettiBottaro2004} in the context of a spatial, rather than temporal, optimization strategy. The existence of a four-cell optimal was not reported. It will be shown below that the two- and the four-cell states are very robust: when they are used as initial conditions for nonlinear simulations the flow trajectory is uncapable of evolving away from them towards a different topology which eventually exploits other symmetries, except when exceedingly large disturbance energies are given as input.

For non-zero streamwise wavenumber, sub-optimal perturbations are found which take the form of modulated travelling waves. For $\alpha=1$, $G=381.41$ at $T=0.384$ and for $\alpha=2$, $G=237.05$ at $T=0.246$. To represent a wave train the temporal dependence of the generic disturbance can be written as the product of an exponential wave-part and an envelope function slowly modulated in time: 
$q(x,y,z,t)=\tilde{q}(y,z,t)exp^{i\alpha(x-ct)}$.  
The phase velocity is found to be quasi-constant with time and close to the bulk velocity: $c(\alpha=1)=1.1117$ and $c(\alpha=2)=1.1536$, both scaled with $U_b$. The study of the temporal (rather than the spatio-temporal) evolution of disturbances is acceptable when flow structures travel at a well-defined speed within the duct. Such an approximation appears to be reasonably well satisfied in experiments on equilibrium puffs in pipe flow (\cite{Hofetal2005}), which are found to be advected downstream at speeds slightly larger than $U_b$, for values of the Reynolds number $Re_b = U_b D / \nu$ ($D$ pipe diameter) exceeding $Re_b \approx 1800$. Below such a threshold, turbulence can no longer be maintained autonomously.

\section{Nonlinear evolution}\label{Nle}

Temporally evolving simulations have been conducted in a periodic duct of length $4\pi$ for a variety of initial conditions at $Re_\tau=150$; such a value is close to the threshold of self-sustained turbulence, as confirmed very recently by \cite{Uhlmann}.  For each direct numerical simulation the initial state consists of the quasi-parabolic base flow profile plus a optimal or sub-optimal perturbation, normalized with prescribed energy $E_0$, plus random noise. The complex amplitude of the noise in Fourier space varies between $\pm10^{-10}$; the noise is necessary to fill the spectrum.

\begin{figure}
\begin{center}
\includegraphics[width=6cm]{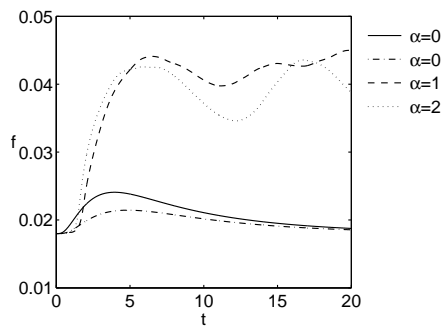}
\caption{\label{skin-friction} Skin friction for the four different optimal perturbations. For $\alpha=0$, the initial energy $E_0$ is equal to $10^{-1}$, for $\alpha=1$, $E_0=7.8\times 10^{-3}$ and for $\alpha=2$, $E_0=4.4\times 10^{-3}$.}
\end{center}
\end{figure}

In figure \ref{skin-friction} the time evolution of the skin friction factor $f$ is plotted for four initial conditions of different initial energies.
When the simulations are initiated with one of the two global optimal solutions of figure \ref{gain} the ensuing behavior is uneventful ({\it cf.} figure \ref{skin-friction}), and even for a rather large initial disturbance amplitude, $E_0 = 0.1$, no instability appears to modify the flow which returns slowly to the laminar condition with $f=0.018$. The results are more suggestive when linear travelling waves are used to initiate the nonlinear computations. For the case in which the condition at $t=0$ is the sub-optimal state with $\alpha = 1$ or $2$, the initial growth of $f$ is slower than for the global optimal case but, by time $t=2$, a strong mean flow deviation is created by nonlinear interactions, leading to a value of the friction factor which oscillates around the turbulent mean value $f = 0.0415$. It is notable that the energies $E_0$ of the sub-optimal initial conditions sufficient to trigger transition are very much lower than $10^{-1}$.

\begin{figure}
\begin{center}
\includegraphics[width=13cm]{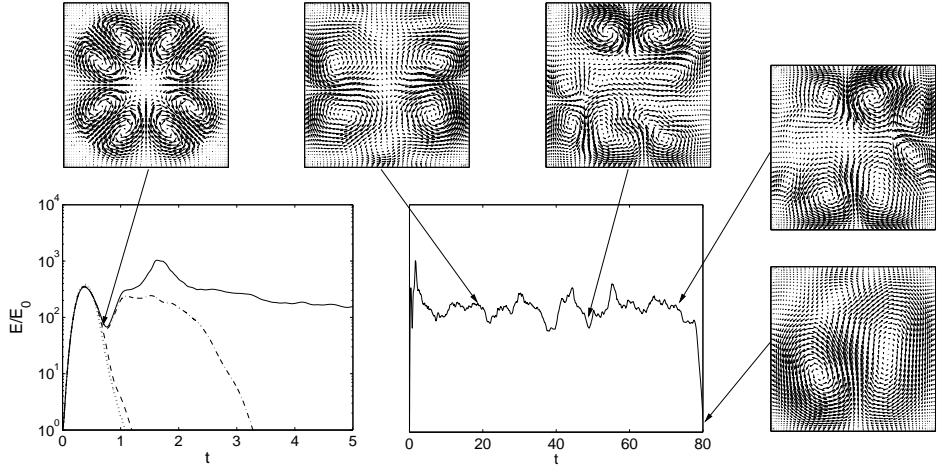}
\caption{\label{seuils}  Evolution in time of the fluctuation energy, initial condition is given by the sub-optimal perturbation with $\alpha=1$. Different curves correspond to different initial energy values: continuous line corresponds to $E_0=7.8\times 10^{-3}$, dash-dotted line to $E_0=7.7\times 10^{-3}$; dashed line to $E_0=5\times 10^{-3}$ and dotted line represents the linear computation. The energy is normalized with its initial value $E_0$. The streamwise averaged vortices are drawn at times $t=0.8,~20,~50,~70$ and $80$.}
\end{center}
\end{figure}

\begin{figure}
\begin{center}
\includegraphics[width=13cm]{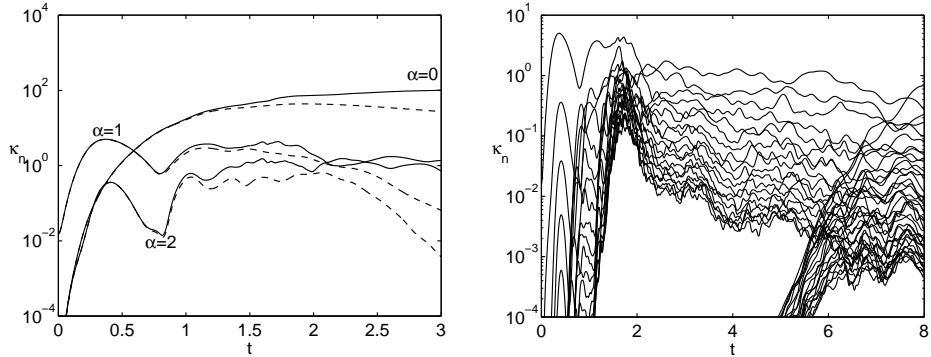}
\caption{\label{spectrum} Left: temporal behavior of three different Fourier modes ($\alpha=0,~1,~2$) for the nonlinear simulation initiated by the optimal $\alpha = 1$ initial condition with $E_0=7.8\times 10^{-3}$ (continuous lines) and $E_0=7.7\times 10^{-3}$ (dashed lines), plus low amplitude noise. On the right the long time evolution of the fluctuations ($\alpha=0.5,~1,~1.5,...$) for the case $E_0=7.8\times 10^{-3}$ is represented. The power density spectrum $\kappa_n$ is defined by $\kappa_n = \frac{1}{N^2}\int_{yz} (\mathbf{\tilde{u}}^*\mathbf{\tilde{u}})_n + (\mathbf{\tilde{u}}^*\mathbf{\tilde{u}})_{N-n}~dydz$, with $\mathbf{\tilde{u}}_n\left(\alpha_n,y,z,t \right)$ the $x$-Fourier transform of $\mathbf{u}$, from which the laminar profile has been substracted.}
\vspace{-.3cm}
\end{center}
\end{figure}

We have explored in more details the long time behaviour of the flow when the sub-optimal initial perturbation with $\alpha=1$ is used to trigger transition; the results are condensed in figure \ref{seuils}. 
As a function of the value of $E_0$ the Navier-Stokes calculations permit to recover the linear behaviour (when $E_0 \leq 5\times 10^{-3}$) or to depart from it. 
The threshold for transition is found for $E_0=7.8\times10^{-3}$;
below such an initial energy value the flow returns rapidly (within a few units of time) to the laminar state, above it the flow becomes turbulent. The energy gain of figure \ref{seuils} shows that when $E_0=7.8\times 10^{-3}$ there is, at $t=0.8$, a sudden increase of the fluctuation energy, likely linked to an instability of the distorted mean flow. There is an interesting connection here with the recent theory of 'minimal defects'\cite{BLC,Biau1,Gavarini,BenDov1}.
The secondary flow at $t=0.8$ shown in the same figure displays symmetries about bisectors and diagonals, but this is not a generic occurrence and different initial conditions generate mean flow defects with other symmetries. For $E_0=7.7\times 10^{-3}$ the growth is followed by decay and rapid relaminarization (\emph{cf.} also figure \ref{spectrum} left); when $E_0=7.8\times 10^{-3}$ the growth which starts at $t=0.8$ is followed by a rapid filling of the spectrum with a peak in the intensity of fluctuations at $t=1.8$. Such a filling is made clear by figure \ref{spectrum} (right): first the modes with $\alpha=n$, $n\in \mathbb{N}$, grow because of quadratic interactions, then the modes with $\alpha=(2n-1)/2$, $n\in \mathbb{N}$, emerge out of the random noise (visible in the figure after $t=5$). Figure \ref{spectrum} (left) focusses on the modes $\alpha=0$ and $\alpha=2$ which are first to be produced by nonlinearities. The streamwise-independent mode remains amplified for a long time because of lift-up effect, its appearance out of the $\alpha=\pm 1$ fundamental mode is the analogue of the so-called \emph{oblique transition} process in channel flow (\cite{oblique}). By time $t=3$ the turbulent flow can be considered as fully developed. Such an edge state persists until $t\approx78$ (\emph{cf.} \ref{seuils}); it remains dynamically connected to the laminar base flow solution since relaminarization is abruptly reached at $t \approx 80$, after coalescence of the smaller vortices into a pair and then into a single large vortex. While turbulence is maintained, the secondary patterns displayed at $t=20,~50$ and $70$ in figure \ref{seuils} resemble the four-cell global optimal disturbance. At $t=20$, two pairs of vortices are clearly visible in the cross-section; they are close to the two vertical walls which can thus be defined as 'active' since it is there that the turbulent wall cycle operates (\cite{Uhlmann}). This 4-cells state oscillates while maintaining remarkable coherence for some twenty units of time. Averaging over $t$ yields the same flow pattern with four regular vortices discovered very recently by \cite{Uhlmann}. Past $t=50$, the active walls shift and the large scale vortices lean on the horizontal surfaces, despite the presence of smaller intermittent features that can be found near the left vertical wall at $t=50$ and the right wall at $t=70$. At $t \approx 80$ the flow relaminarizes; tests performed with different lengths of the computational domain show that the threshold $E_0$ for transition remains unaffected, but relaminarization occurs later both for a shorter computational box ($t \approx 90$ when $L_x = 2 \pi$) and a longer box ($t \approx 150$ when $L_x = 6 \pi$), for the same numerical grid density. Interestingly, for the case of the short box the oscillations of the variables around the mean display enhanced amplitudes, highlighting the
fact that a distorted dynamics could be caused by constraining too much the flow structures.

For a geometrical description of the transition process, we choose the phase subspace spanned by two observables: the Reynolds number (based on bulk velocity) and the energy of the streamwise-averaged flow, noted $E_U=0.5\int_{yz} U^2+V^2+W^2 dydz$, with $(U,~V,~W)$ the velocity vector after streamwise averaging.
\begin{figure}
\begin{center}
\includegraphics[width=14cm]{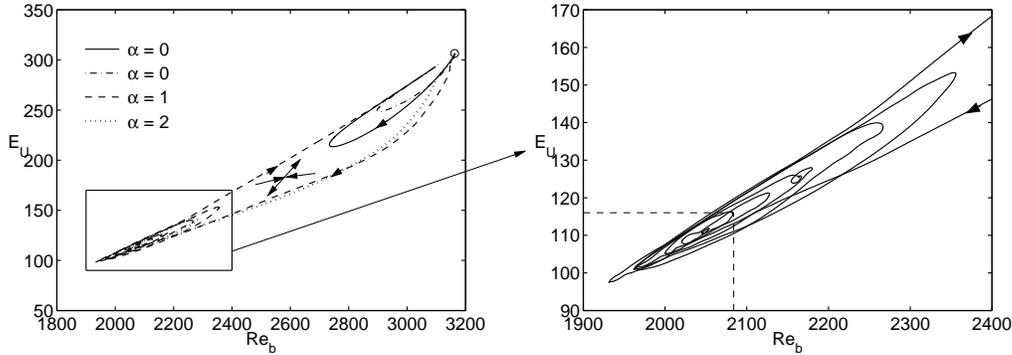}
\caption{\label{PhaseDiag}  
'Phase diagram' representation of the transition process: time evolution in Reynolds number, based on bulk velocity, and mean flow energy $E_U$. The simulations start from the laminar flow solution (open circle in the left figure) plus a (sub-)optimal perturbation. The cross with arrows provides the approximate position of an unstable saddle node. The time averaged values of $E_U$ and $Re_b$ in the fully developed turbulent regime are indicated by dashed lines in the zoom at right which refers only to the initial disturbance perturbations with $\alpha=1$.}
\end{center}
\end{figure}

Figure \ref{PhaseDiag} shows some trajectories between the laminar ($Re_b=3163,~E_U=306.45$) and the turbulent ($Re_b=2084,~E_U=116$) fixed points 
(the latter is simply defined by the temporal average of $E_U$ and $Re_b$ 
in the fully developed turbulent regime). The paths for the initial conditions with $\alpha=0$ correspond to homoclinic orbits in phase space. For the two cases which follow a non-trivial branch ($\alpha=1$ and $2$) the flow approaches the unstable saddle node (qualitatively sketched in the figure) before departing away from it along its unstable manifold. The trajectory in figure \ref{PhaseDiag} (right) then starts circling around the point which characterizes the fully developed turbulent state, with orbits of increasing size.  Before the end of the fifth orbit, it escapes through the unstable manifold of the saddle node towards the laminar fixed point. Each orbit has an ellipsoidal shape and is made up by two portions with very small local radii of curvature, around which the flow spends most of its time, and two long portions of large radii of curvature which are spanned very rapidly by the flow. Relaminarization is consistent with the emerging picture of shear flow turbulence which considers it as a transient event with a characteristic lifetime increasing exponentially with Reynolds number (\cite{nature}).

\section{Discussion and conclusions}\label{sec:concl}
Although non-normality and transient growth are important issues, the traditional emphasis on optimal disturbances may be misleading when trying to predict the onset of transition. The results found here suggest that optimal initial perturbations in the form of steady streamwise vortices play mostly a passive role, while rapidly growing travelling wave packets have the potential to induce transition past a reasonably low threshold value of the initial disturbance energy. This suggests that an energy-like functional is possibly not the most pertinent objective of the optimization procedure. 

In the simulations performed here we have found that key to transition is the establishment of a disturbed mean flow profile (mode $\alpha=0$), susceptible to an exponential or algebraic instability which causes enhanced growth of the travelling wave mode with $\alpha = \pm 1$. Such an early stage can be described by a weakly nonlinear triadic interaction model.

Once turbulence is established, its sustainment depends on non-normality, capable to produce rapid transient growth of the disturbance energy, and non-linearity, which enables directional redistribution of the amplified disturbances. The coupling between transient growth and directional redistribution of energy causes the flow to orbit around the turbulent fixed point in the phase-space of figure \ref{PhaseDiag}.  Near the edge of chaos the lifetime of turbulence is finite and beyond a given value of $t$ (function of the Reynolds number and of the streamwise dimension of the periodic domain) relaminarization occurs. Interestingly, the edge state resembles the optimal disturbance; this fact has recently been observed in pipe flow (\cite{Eckhardt2007,PringleKerswell}) and still awaits for an explanation.

\noindent The financial support of the EU (program Marie Curie EST \textsc{FLUBIO} 20228-2006) and of the Italian Ministry of University and Research (PRIN 2005-092015-002) are gratefully acknowledged.

\bibliographystyle{jfm}
\bibliography{SqDuct}

\end{document}